\documentclass[12pt]{iopart}

\usepackage[english]{babel}
\usepackage[dvips]{graphicx}
\usepackage{amstext,amssymb}
\usepackage[comma,square,numbers,sort]{natbib}
\usepackage{array}
\usepackage{color}

\newcommand{\eqref}[1]{(\ref{#1})}

\newcommand{\eVq}  {\text{eV}^2}

\begin{document}

%% {\hfill IFIC/11-08}

\title[Where we are on $\theta_{13}$: addendum to ``Global neutrino
data...'']{Where we are on $\theta_{13}$: addendum to ``Global neutrino data and recent reactor
  fluxes: status of three-flavour oscillation parameters''}

\author{Thomas Schwetz\dag, Mariam
  T{\'o}rtola\S\ and J. W.~F.~Valle\S}

\address{\dag\ Max-Planck-Institut f\"ur Kernphysik, PO Box 103980, 
69029 Heidelberg, Germany}

\address{\S\ AHEP Group, Instituto de F\'{\i}sica Corpuscular --
  C.S.I.C./Universitat de Val{\`e}ncia, \\
  Edificio Institutos de Paterna, Apt 22085, E--46071 Valencia, Spain}

\ead{schwetz@mpi-hd.mpg.de, mariam@ific.uv.es, valle@ific.uv.es}

\begin{abstract}
  In this addendum to Ref.~\cite{Schwetz:2011qt} we consider the
  recent results from long-baseline $\nu_\mu\to\nu_e$ searches at the
  T2K and MINOS experiments and investigate their implications for the
  mixing angle $\theta_{13}$ and the leptonic Dirac CP phase
  $\delta$. By combining the $2.5\sigma$ indication for a non-zero
  value of $\theta_{13}$ coming from T2K data with global neutrino
  oscillation data we obtain a significance for $\theta_{13} > 0$ of
  about $3\sigma$ with best fit points $\sin^2\theta_{13} = 0.013
  (0.016)$ for normal (inverted) neutrino mass ordering. These results
  depend somewhat on assumptions concerning the analysis of reactor
  neutrino data.
\vskip 1cm
\noindent
keywords: Neutrino mass and mixing; neutrino oscillation; solar and
atmospheric neutrinos; reactor and accelerator neutrinos

\end{abstract}

%%%%%%%%%%%%%%%%%%%%%%%%%%%%%%%%%%%%%%%%%%%%%%%%%%%%%%%%%%%%%%%%%%%%%%%%%%%%%%
\section{Introduction}
%%%%%%%%%%%%%%%%%%%%%%%%%%%%%%%%%%%%%%%%%%%%%%%%%%%%%%%%%%%%%%%%%%%%%%%%%%%%%%

Prompted by the recently published indication for electron neutrino
appearance by the T2K experiment we have updated the global neutrino
oscillation analysis presented in Ref.~\cite{Schwetz:2011qt}. The T2K
experiment uses a neutrino beam consisting mainly of muon neutrinos,
produced at the J-PARC accelerator facility and observed at a distance
of 295~km and an off-axis angle of $2.5^\circ$ by the Super-Kamiokande
detector. The present data release corresponds to $1.43 \times
10^{20}$ protons on target~\cite{Abe:2011sj}. Six events pass all
selection criteria for an electron neutrino event. In a three-flavor
neutrino oscillation scenario with $\theta_{13} = 0$ the expected
number of such events is $1.5\pm0.3$ (syst). Under this hypothesis,
the probability to observe six or more candidate events is $7\times
10^{-3}$, equivalent to a significance of $2.5\sigma$. We investigate
the implications of this result for the mixing angle $\theta_{13}$ and
the leptonic Dirac CP phase $\delta$, focusing on long-baseline
$\nu_\mu\to\nu_e$ appearance data from T2K and MINOS in
sec.~\ref{sec:lbl}, whereas the results of the combined analysis of
global neutrino oscillation data are presented in sec.~\ref{sec:glob}.

%%%%%%%%%%%%%%%%%%%%%%%%%%%%%%%%%%%%%%%%%%%%%%%%%%%%%%%%%%%%%%%%%%%%%%%%%%%%%%
\section{Long-baseline $\nu_\mu\to\nu_e$ appearance data from T2K and MINOS}
%%%%%%%%%%%%%%%%%%%%%%%%%%%%%%%%%%%%%%%%%%%%%%%%%%%%%%%%%%%%%%%%%%%%%%%%%%%%%%
\label{sec:lbl}

For our re-analysis of T2K we use the spectral data shown in Fig.~5 of
Ref.~\cite{Abe:2011sj} given as 5 bins in reconstructed neutrino
energy from 0 to 1.2~GeV. Using the neutrino fluxes predicted at
Super-Kamiokande in the absence of oscillations provided in Fig.~1 of
Ref.~\cite{Abe:2011sj} we calculate the $\nu_\mu\to\nu_e$ appearance
signal by tuning our prediction to the corresponding prediction in
Fig.~5 of Ref.~\cite{Abe:2011sj}. In the fit we include the background
distribution shown in that figure with a systematic normalization
uncertainty of 23\% and adopt the $\chi^2$ definition based on the
Poisson distribution. The calculation is performed by using the GLoBES
simulation software~\cite{Huber:2007ji}.
Latest MINOS data on the $\nu_\mu\to\nu_e$ channel have been presented
in Ref.~\cite{minos11}, corresponding to $8.2 \times 10^{20}$ protons
on target, compared to $7\times 10^{20}$ used in
Ref.~\cite{Schwetz:2011qt}. MINOS finds 62 events with an expectation
in absence of oscillations of $49.6\pm 7.0\text{(stat)}\pm
2.7\text{(syst)}$, showing no significant indication for
$\nu_\mu\to\nu_e$ transitions.

\begin{figure} \centering
\includegraphics[width=0.7\textwidth]{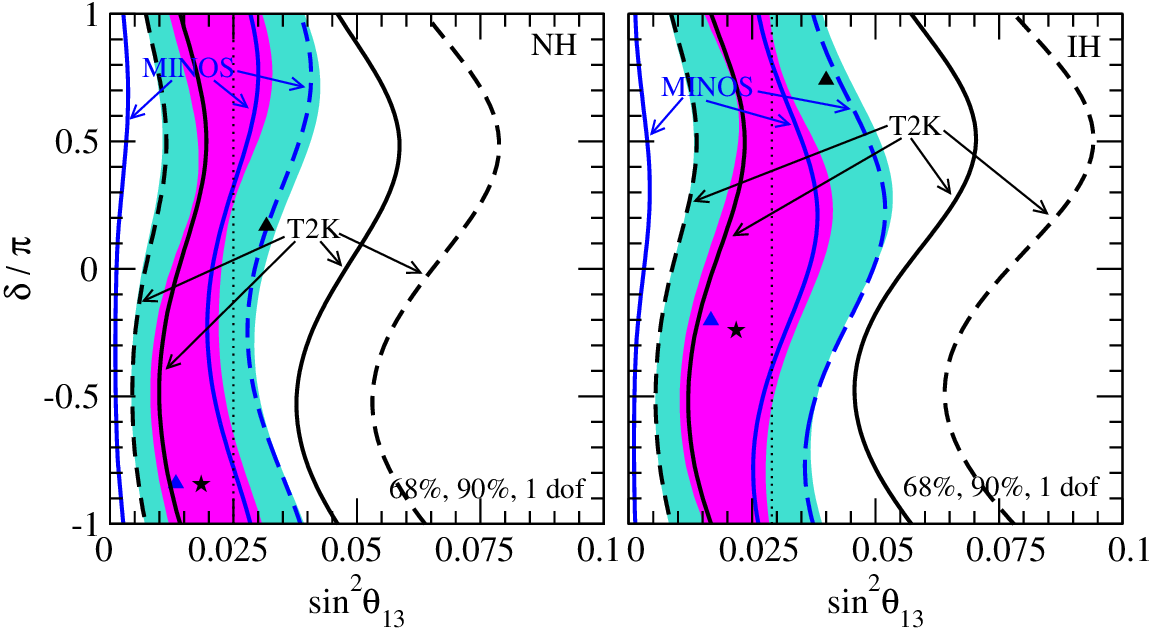} 
\caption{Regions in the $\sin^2\theta_{13} - \delta$ plane at 68\% and
  90\%~CL for 1~dof for T2K and MINOS appearance data (curves) and
  their combination (shaded regions). For all other oscillation
  parameters we assume best fit values and uncertainties according to
  Tab.~\ref{tab:summary}, and we include a 5\% uncertainty on the
  matter density. The left (right) panel is for normal (inverted) mass
  hierarchy. The dotted line shows the 90\%~CL upper limit on
  $\sin^2\theta_{13}$ from a combined analysis of all other
  oscillation data.} \label{fig:LBL}
\end{figure}

In Fig.~\ref{fig:LBL} we show the region in the $\sin^2\theta_{13} -
\delta$ plane indicated by T2K data in comparison to MINOS results.
While for T2K we obtain a closed region for $\sin^2\theta_{13}$ at
90\%~CL ($\Delta\chi^2 = 2.7$), for MINOS we find only an upper
bound. The results are clearly compatible and we show the combined
analysis as shaded regions, where the upper bound is determined by the
MINOS constraint while the lower bound is given by T2K. Best fit
values are in the range $\sin^2\theta_{13} \approx 0.015-0.023$,
depending on the CP phase $\delta$, where the variation is somewhat
larger for the inverted mass hierarchy. The dotted lines in the figure
indicate the 90\%~CL upper bound on $\sin^2\theta_{13}$ coming from a
combined analysis of the remaining oscillation data, including global
reactor, solar, atmospheric, and long-baseline disappearance data.

%%%%%%%%%%%%%%%%%%%%%%%%%%%%%%%%%%%%%%%%%%%%%%%%%%%%%%%%%%%%%%%%%%%%%%%%%%%%%%
\section{Global analysis}
%%%%%%%%%%%%%%%%%%%%%%%%%%%%%%%%%%%%%%%%%%%%%%%%%%%%%%%%%%%%%%%%%%%%%%%%%%%%%%
\label{sec:glob}

\begin{figure}
  \centering
  \includegraphics[width=0.47\textwidth]{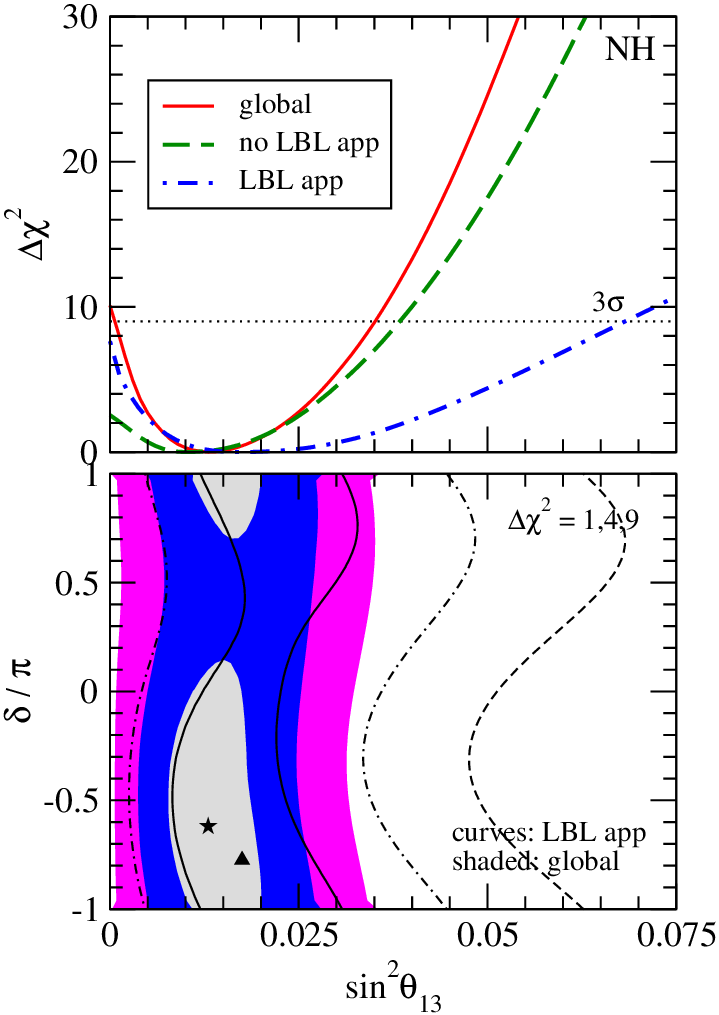}
  \includegraphics[width=0.47\textwidth]{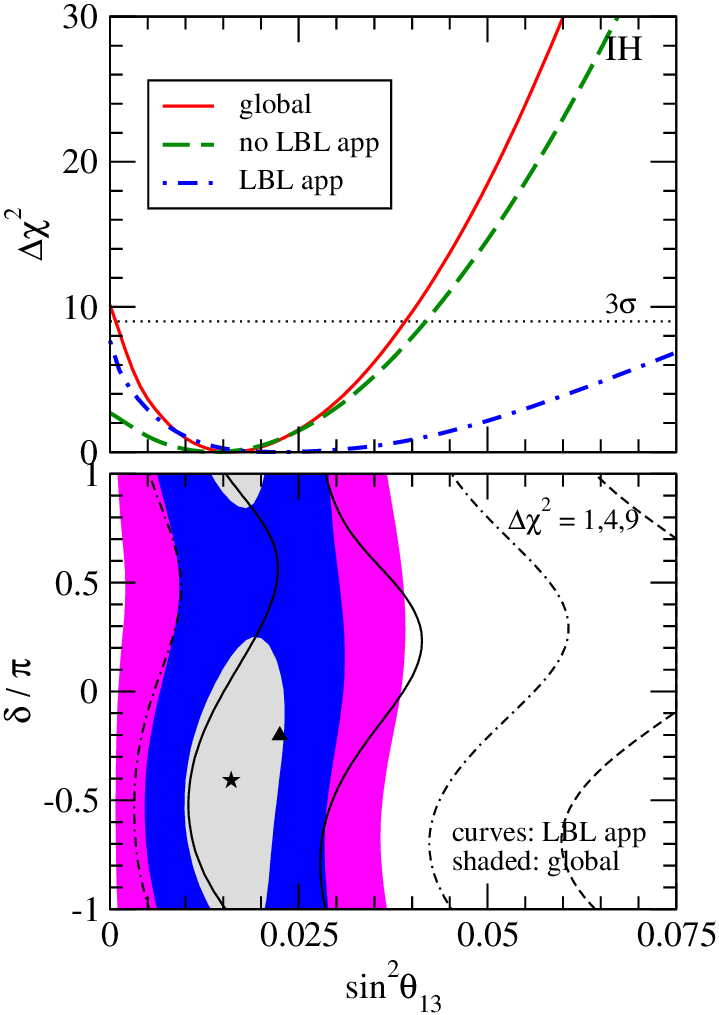}
  \caption{Upper panels: $\Delta\chi^2$ as a function of
    $\sin^2\theta_{13}$ for T2K and MINOS $\nu_e$ appearance data
    (``LBL app''), all the other global data (``no LBL app''), and the
    combined global data (``global'').  Lower panels: contours of
    $\Delta\chi^2=1,4,9$ in the $\sin^2\theta_{13}-\delta$ plane for
    ``LBL app'' (curves) and for the global data (shaded regions). We
    minimize over all undisplayed oscillation parameters. Left (right)
    panels are for normal (inverted) neutrino mass hierarchy.}
\label{fig:glob}
\end{figure}

We move now to the combined analysis of the T2K and MINOS $\nu_e$
appearance searches with global neutrino oscillation data as described
and referenced in Ref.~\cite{Schwetz:2011qt}. For the reactor analysis
we use the ``recommended'' analysis from Ref.~\cite{Schwetz:2011qt},
which adopts the new reactor neutrino fluxes from
Ref.~\cite{Mueller:2011nm} while including short-baseline reactor
neutrino experiments with baselines $\lesssim 100$~km in the fit. The
results for $\theta_{13}$ are summarized in Fig.~\ref{fig:glob}. For
both neutrino mass hierarchies we find that the 2.5$\sigma$ indication
for $\theta_{13} > 0$ from T2K gets pushed to the $3\sigma$ level
($\Delta\chi^2 = 9$) when combined with the weak hint for a non-zero
$\theta_{13}$ obtained from the remaining data \cite{Schwetz:2011qt},
see also Ref.~\cite{Fogli:2011qn}. We find best fit points at
\begin{equation}
\begin{array}{c@{\qquad}l}
\sin^2\theta_{13} = 0.013\,,\quad \delta = -0.61\pi & \text{(normal hierarchy),} \\
\sin^2\theta_{13} = 0.016\,,\quad \delta = -0.41\pi & \text{(inverted hierarchy).} 
\end{array}
\end{equation}

Due to some complementarity between T2K and MINOS one obtains, after
combining with the $\theta_{13}$ limit from the rest of the data, a
``preferred region'' for the CP phase $\delta$ at $\Delta\chi^2=1$, as
seen in Fig.~\ref{fig:glob}.
Obviously this preference for the CP phase is not
significant.\footnote{Prospects to constrain $\delta$ with the
present generation of experiments have been discussed in
Ref.~\cite{Huber:2009cw}.} Marginalizing over the CP phase $\delta$
(and all other oscillation parameters) we obtain for the best fit,
one-sigma errors, and the significance for $\theta_{13} > 0$:
\begin{equation}
\begin{array}{c@{\qquad}l}
\sin^2\theta_{13} = 0.013^{+0.007}_{-0.005}\,,\quad \Delta\chi^2 = 10.1 \,(3.2\sigma) & \text{(normal),} \\
\sin^2\theta_{13} = 0.016^{+0.008}_{-0.006}\,,\quad \Delta\chi^2 = 10.1 \,(3.2\sigma) & \text{(inverted).} 
\end{array}
\end{equation}
As expected the upper bound on $\sin^2\theta_{13}$ is dominated by
global data without long-baseline appearance data, whereas the lower
bound comes mainly from T2K.

Let us briefly consider the sensitivity of these results to the
assumptions on the analysis of reactor neutrino data. As discussed in
detail in Ref.~\cite{Schwetz:2011qt} there is a slight tension between
reactor neutrino fluxes obtained in Ref.~\cite{Mueller:2011nm} and the
results of short-baseline reactor neutrino experiments with baselines
$\lesssim 100$~km.  The increase of reactor neutrino fluxes compared
to previous calculations found in Ref.~\cite{Mueller:2011nm} has been
confirmed qualitatively by an independent recent
calculation~\cite{patrick}. To illustrate the impact on $\theta_{13}$
we show the results for two alternative assumptions for the reactor
analysis. Adopting the fluxes from Ref.~\cite{Mueller:2011nm} but
omitting reactor experiments with baselines $\lesssim 100$~km we find
for the best fit, one-sigma errors, and the significance for
$\theta_{13} > 0$:
\begin{equation}
\begin{array}{c@{\quad}l}
\sin^2\theta_{13} = 0.022 \pm 0.008\,,\quad \Delta\chi^2 = 13.5\,(3.7\sigma) & \text{(NH)} \\
\sin^2\theta_{13} = 0.026 \pm 0.009\,,\quad \Delta\chi^2 = 15.2 \,(3.9\sigma) & \text{(IH)} 
\end{array}
\quad\text{(no SBL react)}
\end{equation}
If instead we do include the short-baseline reactor data but leave the overall normalization of the reactor neutrino flux free we obtain
\begin{equation}
\begin{array}{c@{\quad}l}
\sin^2\theta_{13} = 0.011^{+0.007}_{-0.004}\,,\quad \Delta\chi^2 = 7.7 \,(2.8\sigma) & \text{(NH)} \\
\sin^2\theta_{13} = 0.014^{+0.007}_{-0.006}\,,\quad \Delta\chi^2 = 8.4 \,(2.9\sigma) & \text{(IH)} 
\end{array}
\quad\text{(flux free)}
\end{equation}
We see that the precise value of the $\sin^2\theta_{13}$ best fit
point as well as the significance for $\theta_{13} > 0$ still depend
on assumptions on the reactor analysis, as discussed in detail in
Ref.~\cite{Schwetz:2011qt}.

\begin{table}[ht]\centering
   \catcode`?=\active \def?{\hphantom{0}}
   \begin{tabular}{|@{\quad}>{\rule[-2mm]{0pt}{6mm}}l@{\quad}|@{\quad}c@{\quad}|@{\quad}c@{\quad}|@{\quad}c@{\quad}|}
       \hline
       parameter & best fit $\pm 1\sigma$ & 2$\sigma$ & 3$\sigma$
       \\
       \hline
       $\Delta m^2_{21}\: [10^{-5}\eVq]$
& $7.59^{+0.20}_{-0.18}$ & 7.24--7.99 & 7.09--8.19 \\[3mm] %%
       $\Delta m^2_{31}\: [10^{-3}\eVq]$
&
   \begin{tabular}{c}
     $2.50^{+0.09}_{-0.16}$\\
     $-(2.40^{+0.08}_{-0.09})$
   \end{tabular}
&
   \begin{tabular}{c}
     $2.25-2.68$\\
     $-(2.23-2.58)$
   \end{tabular}
&
   \begin{tabular}{c}
     $2.14-2.76$\\
     $-(2.13-2.67)$
   \end{tabular}
   \\[6mm] %% igual
       $\sin^2\theta_{12}$
& $0.312^{+0.017}_{-0.015}$ & 0.28--0.35 & 0.27--0.36\\[3mm]  %%
       $\sin^2\theta_{23}$
&
   \begin{tabular}{c}
     $0.52^{+0.06}_{-0.07}$\\
     $0.52\pm0.06$
   \end{tabular}
&
   \begin{tabular}{c}
         0.41--0.61\\
         0.42--0.61
       \end{tabular}
       & 0.39--0.64 \\[5mm] 
       $\sin^2\theta_{13}$
&
   \begin{tabular}{c}
         $0.013^{+0.007}_{-0.005}$\\
     $0.016^{+0.008}_{-0.006}$
   \end{tabular}
&
   \begin{tabular}{c}
    0.004--0.028\\
        0.005--0.031
   \end{tabular}
&
   \begin{tabular}{c}
         0.001--0.035\\ %%
         0.001--0.039
   \end{tabular}\\[5mm] %%igual
       $\delta$
&
   \begin{tabular}{c}
   %  $[-\pi, 0.14\pi] \& [0.70\pi, \pi]$ or
         $\left(-0.61^{+0.75}_{-0.65}\right)\pi$\\
       %  $[-\pi, 0.25\pi] \& [0.84\pi, \pi]$ or
         $\left(-0.41^{+0.65}_{-0.70}\right)\pi $
   \end{tabular}
&
   $0-2\pi$
&
   $0-2\pi$ \\
       \hline
\end{tabular}
\caption{ \label{tab:summary} Neutrino oscillation parameters
  summary. For $\Delta m^2_{31}$, $\sin^2\theta_{23}$,
  $\sin^2\theta_{13}$, and $\delta$ the upper (lower) row corresponds to normal (inverted)
  neutrino mass hierarchy. See Ref.~\cite{Schwetz:2011qt} for details and references.}
\end{table}

To summarize we display the status for all neutrino oscillation
parameters from the global analysis using the default reactor
treatment in Tab.~\ref{tab:summary}.
If $\theta_{13}$ is indeed within the presently indicated range we may
expect a confirmation by future T2K data soon. Depending on whether
its true value is close to the upper of lower edge of the presently
favored range, an independent confirmation of a non-zero $\theta_{13}$
may be expected from reactor experiments within few months to few
years \cite{Mezzetto:2010zi}.
After establishing the LMA-MSW solution to the solar neutrino problem,
the present $3\sigma$ indication for a non-zero $\theta_{13}$ may turn
out to be first sign for the second necessary ingredient for
observable CP violation in neutrino oscillations, see
e.g.~\cite{Nunokawa:2007qh} for a review.

\section*{Acknowledgments}

Work supported by Spanish grants FPA2008-00319/FPA, MULTIDARK
Consolider CSD2009-00064, PROMETEO/2009/091, and by EU network UNILHC,
PITN-GA-2009-237920. M.T.\ acknowledges financial support from CSIC
under the JAE-Doc programme. This work was partly supported by the
Transregio Sonderforschungsbereich TR27 ``Neutrinos and Beyond'' der
Deutschen Forschungsgemeinschaft.

\bibliographystyle{unsrt}

\end{document}